\documentstyle[pmart,twoside,fleqn,epsf]{article} 

\begin{document}
\newcommand{\labs} {\left\vert}
\newcommand{\rabs} {\right\vert}
\newcommand{\lsb} {\left[}
\newcommand{\rsb} {\right]}
\newcommand{\lrb} {\left(}
\newcommand{\rrb} {\right)}
\newcommand{\lcb} {\left\{}
\newcommand{\rcb} {\right\}}
\newcommand{\lab} {\left\langle}
\newcommand{\rab} {\right\rangle}
\newcommand{\ve} {\varepsilon}
\newcommand{\com}{\%}
\newcommand{\var}{\rho}

%
%
\title{Comparative Studies on Giant Magnetoresistance in Carbon
Nanotubes and Graphene Nanoribbons with Ferromagnetic Contacts}

\author{S. Krompiewski$^{a}$}

%
%
\address{$^{a}$Institute of Molecular Physics, Polish Academy of
  Sciences \\  M. Smoluchowskiego 17, 60-179 Pozna\'n, Poland
}
%
%

\date{May 8, 2008}
\maketitle                   
%
%
\begin{abstract}
 This contribution reports on comparative studies on giant
 magnetoresistance (GMR) in carbon nanotubes (CNTs) and graphene nanoribbons of similar
aspect ratios (i.e perimeter/length and width/length ratios, for
the former and the latter, respectively). The problem is solved at
zero temperature in the ballistic transport regime, by means of
the Green's functions technique within the tight-binding model and
with the so-called wide band approximation for electrodes. The GMR
effect in graphene is comparable to that of CNTs, it depends
strongly on the chirality and only slightly on the aspect ratio.
It turns out that graphene, analogously to CNTs may be quite an
interesting material for spintronic applications.
\end{abstract}

%
%
\section{Introduction}

      Carbon-based structures evoke enormous interest in search for new
materials for future nanoelectronics, expected to replace the
conventional electronics soon. Along with carbon nanotubes
\cite{Saito}, also graphene has recently given an additional
impetus to this type of studies, after it was demonstrated that
individual monolayers of graphite can be successfully fabricated
and electrically contacted \cite{novoselov}.
 Although it has already been realized for several years
 that ferromagnetically contacted carbon nanotubes reveal quite a
noticeable giant magnetoresistance (GMR) effect (\cite{Tsuka}
-\cite{ustron}),
in the case of grephene this problem still remains
to be explored.

The paper is organized as follows: Sec. 2 presents the model and
the adopted method of calculations, in Sec. 3 the results are
presented and discussed. Finally, Sec.~4 summarizes the main
conclusions.

\section{Methodology}

A single-band tight-binding model for non-interacting
$\pi$-electrons is used both for CNTs and graphene (Gr):

\begin{equation} \label{H}
H =\sum \limits_{ i , \sigma } \varepsilon_{i}\labs i, \sigma \rab
\lab \sigma, i \rabs+ \sum \limits_{ i ,j, \sigma } t_{i,j}\labs
i, \sigma \rab \lab \sigma, j \rabs,
\end{equation}

In the following, the nearest neighbor hopping integral (typically
equal to 2.7 - 3 eV) will be treated as the energy unit, t=1,
whereas the on-site energies $\varepsilon$ depend on a gate
voltage.
 The  Hamiltonian~[\ref{H}] can be written in a tri-diagonal block form,
 each block (sub-matrix) is of rank
 equal to the number of atoms within the unit cell (N). For the armchair-CNT
 (ac-CNT) and zigzag-edge graphene (zz-Gr), as well as for the
 zigzag-CNT (zz-CNT) and armchair graphene (ac-Gr)
 structures discussed here $N = 4n$, where $n$ is the
 component of the respective chiral vector $(n, n)$, $(n, 0)$ for the ac-CNT
 and the zz-CNT.
 The unit cells are repeated periodically M times in the
 charge transport direction, resulting in 4nM atoms in the system.
 The present recursive method makes it possible, in principle, to
 perform recursive computations even for systems approaching
 realistic lengths of several hundred nm \cite{KMB}. This time however relatively
 short systems are dealt with, because the systems of interest
 include also those with large transverse dimensions (exceeding the
 longitudinal ones).

As regards ferromagnetic leads (electrodes), the so called
wide-band approximation has been used, i.e. the surface Green's
functions of the leads are treated as energy-independent, but
spin-dependent (i.e. different for $\uparrow$- and
$\downarrow$-spin electrons). The recursive method (see
\cite{Lake}, \cite{KMB}]) is
 described by the following set of equations:

\begin{equation} \label{g}
g_{i,i}^L=(E-D_i-\Sigma_{i}^L)^{-1}, \; \; \;
g_{i,i}^R=(E-D_i-\Sigma_{i}^R)^{-1},
\end{equation}

\begin{equation} \label{sigmas}
\Sigma_{i}^L=T_{i,i-1} g_{i-1,i-1}^R T_{i-1,i}, \; \; \;
\Sigma_{i}^R=T_{i,i+1} g_{i+1,i+1}^R T_{i+1,i},
\end{equation}

\begin{equation} \label{G}
G_{i}=(E-D_i-\Sigma_{i}^L-\Sigma_{i}^R)^{-1}.
\end{equation}

Eqs.~(\ref{g}) define Green's functions for the i-th unit cell,
the matrices D and T stand for the diagonal and off--diagonal
sub-matrices, whereas the full Green's function is given by
Eq.~\ref{G}. The carbon-based structure (either CNT or Gr) extends
from i=1 up to i=M, whereas the left (L) and right (R) leads have
indices $i < 1$ and $i >M$, respectively. The simplest
approximation one can think of, for the leads is the
aforementioned wideband approximation. When applied to the present
situation it corresponds to the following substitutions of the
surface self-energies $\Sigma$'s by purely imaginary,
energy-independent diagonal matrices:

\begin{equation} \label{Sigma}
\Sigma_{0}^L=- j \hat 1 \Delta_L/2 , \; \; \; \Sigma_{M+1}^R=-j
\hat 1 \Delta_R/2  \; \; \; \;( {\rm with} \; j =\sqrt{-1}).
\end{equation}

The other quantities of the main interest here are transmission
$\cal T$, conductance ($\cal G$) and giant magnetoresistance
(GMR). In the ballistic transport regime and at zero temperature
these quantities read:

\begin{equation} \label{T}
{\cal T}=Tr(\Gamma_{i}^L G_{i} \Gamma_{i}^R G_{i}^{\dagger}),
\;\;\Gamma_{i}^{L,R}=j \lsb \Sigma_{i}^{L,R}-\lrb
\Sigma_{i}^{L,R}\rrb ^\dagger \rsb,
\end{equation}

\begin{equation} \label{GMR}
 GMR=100 (1-{\cal G}_{\uparrow,\downarrow}/{\cal
 G}_{\uparrow,\uparrow}), \; \; \; {\cal G} = \frac{e^2}{h} {\cal T}(E_F),
\end{equation}

where the arrows $\uparrow \uparrow$ and $\uparrow \downarrow$
denote prallel and anriparallel alignments of ferromagnetic
electrodes. It is noteworthy that the transmission does not depend
on the reference unit number $i$. The Fermi energy $E_F=0$ at zero
gate voltage ($V_G$) and may be shifted in energy by $\alpha V_G$
for a finite gate voltage, where $\alpha$ is a conversion factor.

\section{Results and discussion}

\begin{figure}[h]
  \epsfysize=3.5cm           
  \centerline{\epsfbox{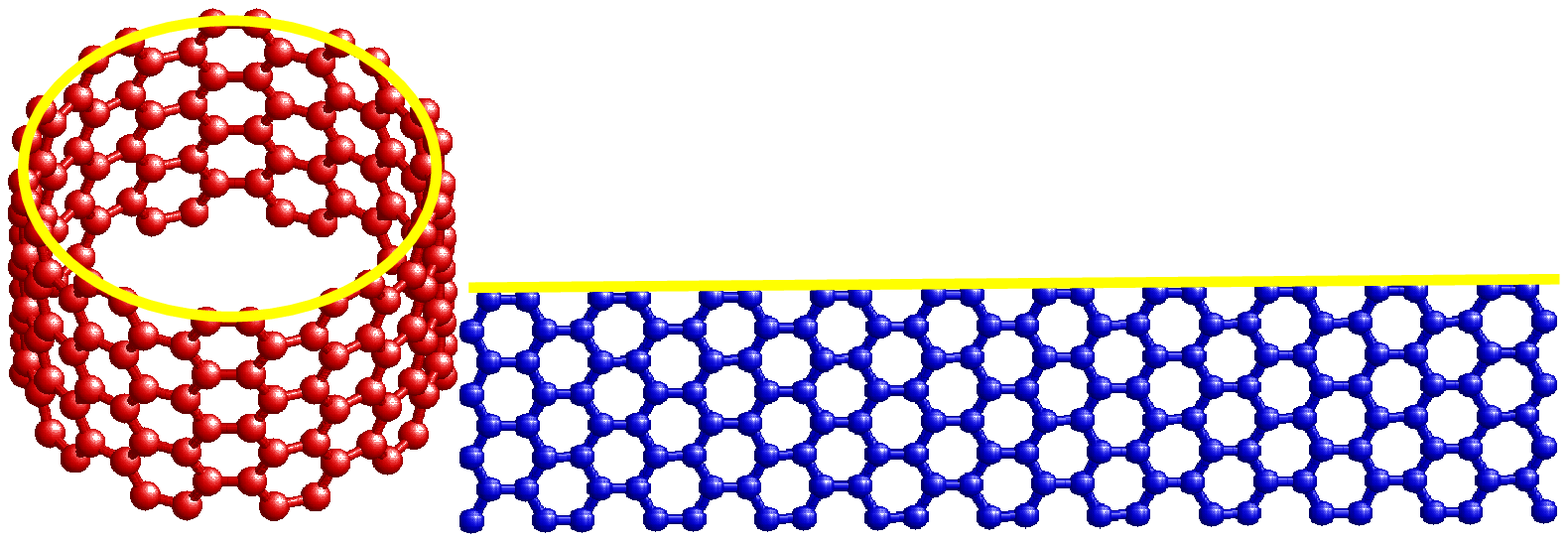}}    
  \caption{Example of a CNT and a graphene sheet of similar aspect ratios.
  The leads are to be attached in the vertical direction both to the CNT and graphene.}
  \label{balls}
\end{figure}

The systems of interest here are CNTs and graphene ribbons of the
same chirality, and similar aspect ratios defined as A=W/L, where
L stands for the length and W for the width in the case of
graphene, and the perimeter in the case of the CNT. An exemplary
illustration of such systems is presented in Fig.~\ref{balls}.

\begin{figure}[ht]
\epsfxsize=10cm                          
  \centerline{\epsfbox{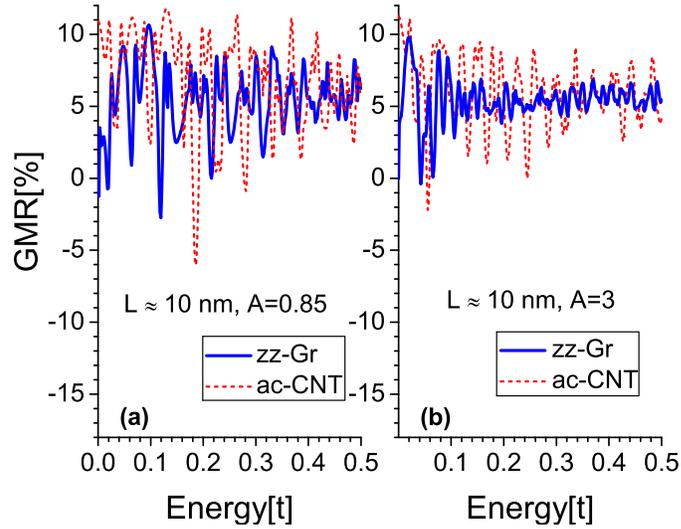}}    
  \caption{GMR for ac-CNT (dashed line) and zz-graphene (solid line), and two
  different aspect ratios: less than 1 (a) and greater than 1 (b).}
  \label{zzGr}
\end{figure}

\begin{figure}[!htb]
  \epsfxsize=10cm           
  \centerline{\epsfbox{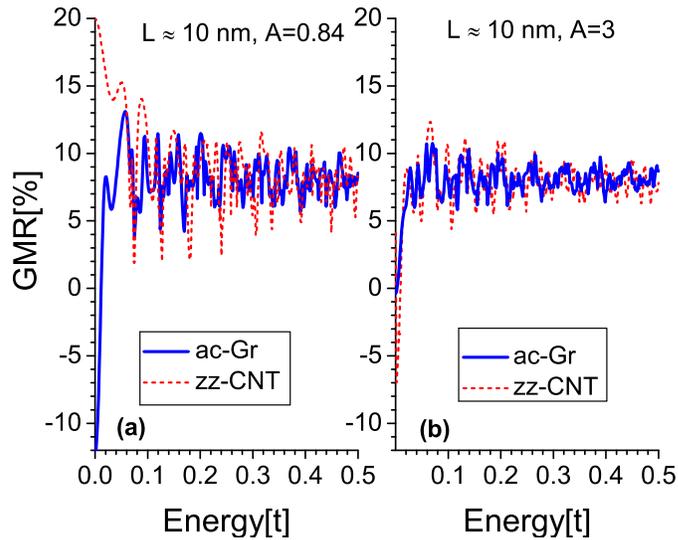}}    
  \caption{As Fig.~\ref{zzGr} but for different chiralities, i.e zz-CNT and ac-Gr.}
  \label{acGr}
\end{figure}

In the case of ferromagnetic leads, the parameters $\Delta$ have
been chosen as follows $\Delta_{L,R}^\sigma=\Delta_0 (1 \pm \sigma
m)$, with $\Delta_0=0.6$, $m=1/3$. If the ferromagnetic leads are
parallel aligned, the plus sign should be taken for both the sides
(L and R), otherwise (antiparallel alignment) - the plus-sign
applies to the L-lead and the minus-sign to the R-lead. The
following two figures show how the GMR of CNTs and Gr sheets
compare with each other for different aspect ratios (\emph{cf.}
\emph{(a)} and \emph{(b)}) and different chiralities (\emph{cf.}
Figs.~\ref{zzGr}, \ref{acGr}).

It is readily seen from these plots that the aspect ratio has some
effect on the GMR. In particular, the long and narrow CNT show
quasi-periodic behavior within the energy region below the onset
to the higher sub-band, clearly revealing 2 and a half periods
(see the first three narrow minima of the dashed line in
Fig.~\ref{zzGr}a). The quasi-period is known to amount to $\hbar
v_F/(2L)$. No distinct periodicity is seen in the case of
graphene, where the longitudinal interference patterns are
affected by the transverse modes and less pronounced Fabry-Perot
patterns can be observe \cite{miao}. As shown in Fig.~\ref{zzGr}b
the periodicity is lost even for the ac-CNT if the aspect ratio is
high enough. It is so, since for $A>1$ the energy inter-level
spacings ($\sim 1/L$) becomes comparable with the distance between
the energy sub-bands ($\sim 1/W$).

As concerns the GMR, in the case of small $A$ negative values
(inverse GMR) can be acquired, as already predicted theoretically
for a more realistic model of electrodes \cite{KGC} and observed
experimentally \cite{schoe}. This tendency may be explained basing
on the Wigner-Breit formula for the on-resonance case and some
anisotropy in CNT/electrode couplings. Interestingly for the
present parametrization the average value of GMR changes roughly
from 5\% up to 10\% for the zz-Gr, ac-CNT and ac-Gr, zz-CNT,
respectively, i.e depends strongly on the chirality of the
structure at hand, but only slightly on the aspect ratio. Nearby
the charge neutrality point, $E_F=0$, the GMR behaves differently
for the CNTs and graphene ribbons. This point is however very
peculiar in many respects (massless fermions, minimum conductivity
effect, non-negligible spin orbit coupling, etc.) and requires
more sophisticated approach than the present one. It should be
noticed in this context that here the spin-orbit (SO) coupling has
been neglected. In the light of recent experiments \cite{kuemmeth}
this is definitely justified in the case of graphene, but much
less in the case of CNTs. Nevertheless, there is a consensus that
SO coupling is not large enough to completely destroy the spin
coherence (especially in short systems like those considered
here).

\section{Conclusion}

Graphene ribbons show the GMR effect of comparable magnitude to
CNTs with similar aspect ratios. The most important findings of
the present studies are: \emph{(i)} the mean value of GMR is
bigger in ac-Gr than in zz-Gr, \emph{(ii)} the fluctuations of GMR
decrease with the increasing aspect ratio $A$, and \emph{(iii)}
for graphene the GMR fluctuations, as a function of energy (gate
voltage), are smaller than those of its CNT counterpart.

Summarizing, both the carbon-based structures under consideration
here seem to be quite promising for potential spintronic
applications.

\section*{Acknowledgments}
This work was suported by the EU FP6 grants: CARDEQ under contract
No. IST-021285-2, and SPINTRA under contract No.
ERAS-CT-2003-980409.

\end{document}